\documentclass[useAMS,usenatbib]{mn2e}
 
\usepackage{graphicx}
\usepackage{graphics,float}                
\usepackage{epsf}
 
\usepackage{amsmath}                                
\usepackage{amssymb}
\usepackage{picinpar}                               
\usepackage{relsize,setspace,subfigure}
\usepackage{tabularx}
\usepackage[innercaption]{sidecap}                  
\usepackage[figuresright]{rotating}
\usepackage{url}
 
\usepackage{natbib} 
 
\def\etal{{\it et al}.}

\def\d3k{{\displaystyle {\rm d}{\bf k} \over \displaystyle (2\pi)^3}}

\title[Alignment of Voids in the Cosmic Web]{Alignment of Voids in the Cosmic Web}
 
\author[Platen, van de Weygaert \& Jones]{Erwin Platen\thanks{E-mail: platen@astro.rug.nl (EP)},
Rien van de Weygaert \& Bernard J.T. Jones$^{1}$\\
  $^{1}$Kapteyn Astronomical Institute, University of Groningen, P.O.
  Box 800, 9700 AV, Groningen, The Netherlands.}
 
\begin{document}
 
\date{Accepted .... Received ...; in original form ...}
 
\pagerange{\pageref{firstpage}--\pageref{lastpage}} \pubyear{2007}
 
\maketitle
 
\label{firstpage}
 
\begin{abstract}
We investigate the shapes and mutual alignment of voids in the large
scale matter distribution of a $\Lambda$CDM cosmology simulation.  The
voids are identified using the novel WVF void finder technique.  The
identified voids are quite nonspherical and slightly prolate, with
axis ratios in the order of $c:b:a\approx 0.5:0.7:1$. Their
orientations are strongly correlated with significant alignments
spanning scales $>30h^{-1}\hbox{Mpc}$.
 
We also find an intimate link between the cosmic tidal field and the
void orientations. Over a very wide range of scales we find a coherent
and strong alignment of the voids with the tidal field computed from
the smoothed density distribution. This orientation-tide alignment
remains significant on scales exceeding twice the typical void size,
which shows that the long range external field is responsible for the
alignment of the voids.  This confirms the view that the large scale
tidal force field is the main agent for the large scale spatial
organization of the Cosmic Web.
\end{abstract}
 
 
\section{Introduction}
Galaxy redshift surveys and computer simulations of cosmic structure
formation have shown that matter traces out a frothy pattern, the
Cosmic Web.  \cite{bondweb1996} established how the tidal forces
induced by the inhomogeneous cosmic matter distribution are the main
agent shaping the Cosmic Web.  The resulting cosmic tidal force field
is the source of the large scale, coherent, anisotropic forces which
shape the cosmic matter distribution into characteristic filamentary
and sheetlike patterns.
 
\begin{figure*}
  \includegraphics[width=0.45\textwidth]{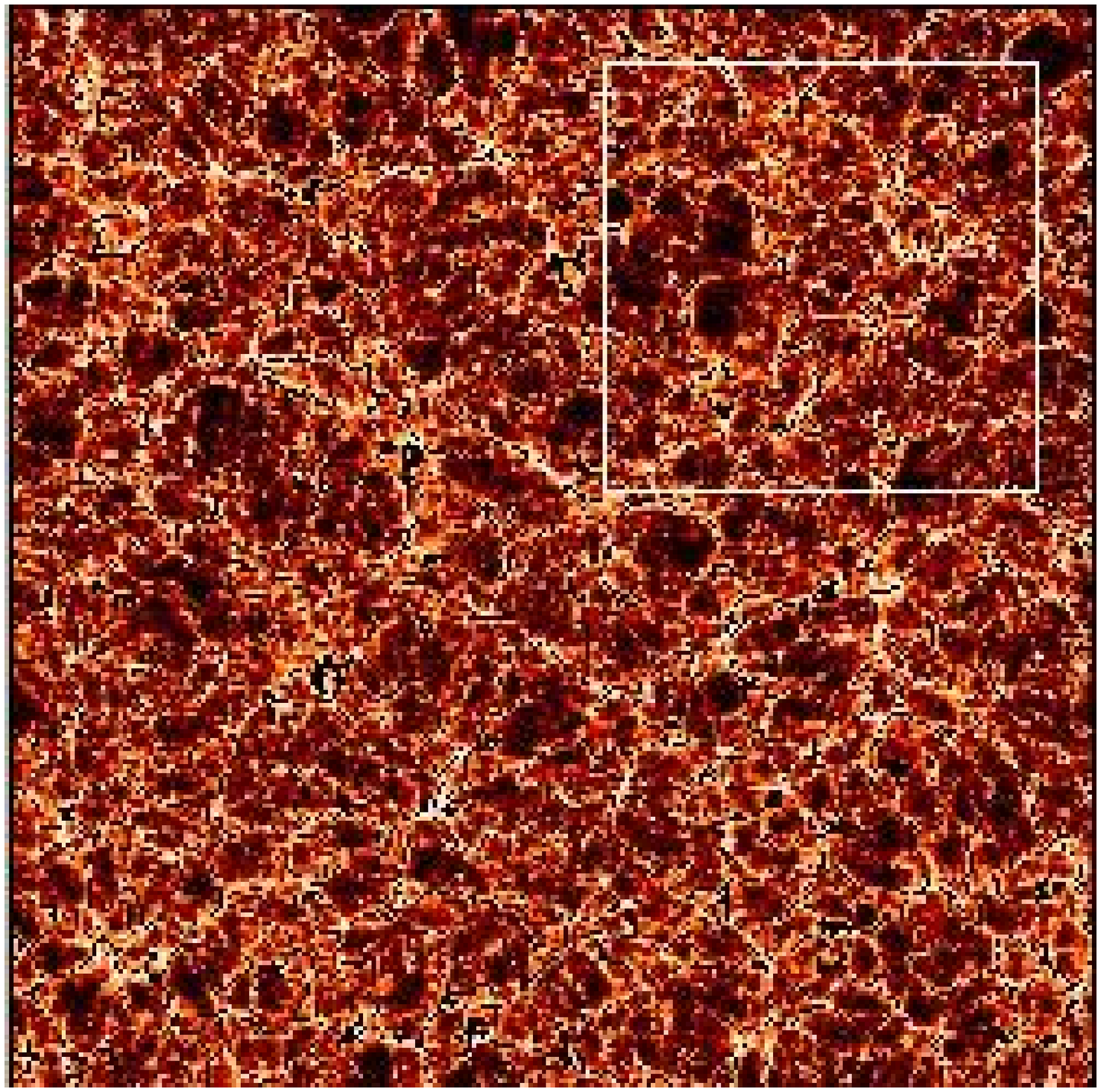}\hspace{0.5cm}
  \includegraphics[width=0.45\textwidth]{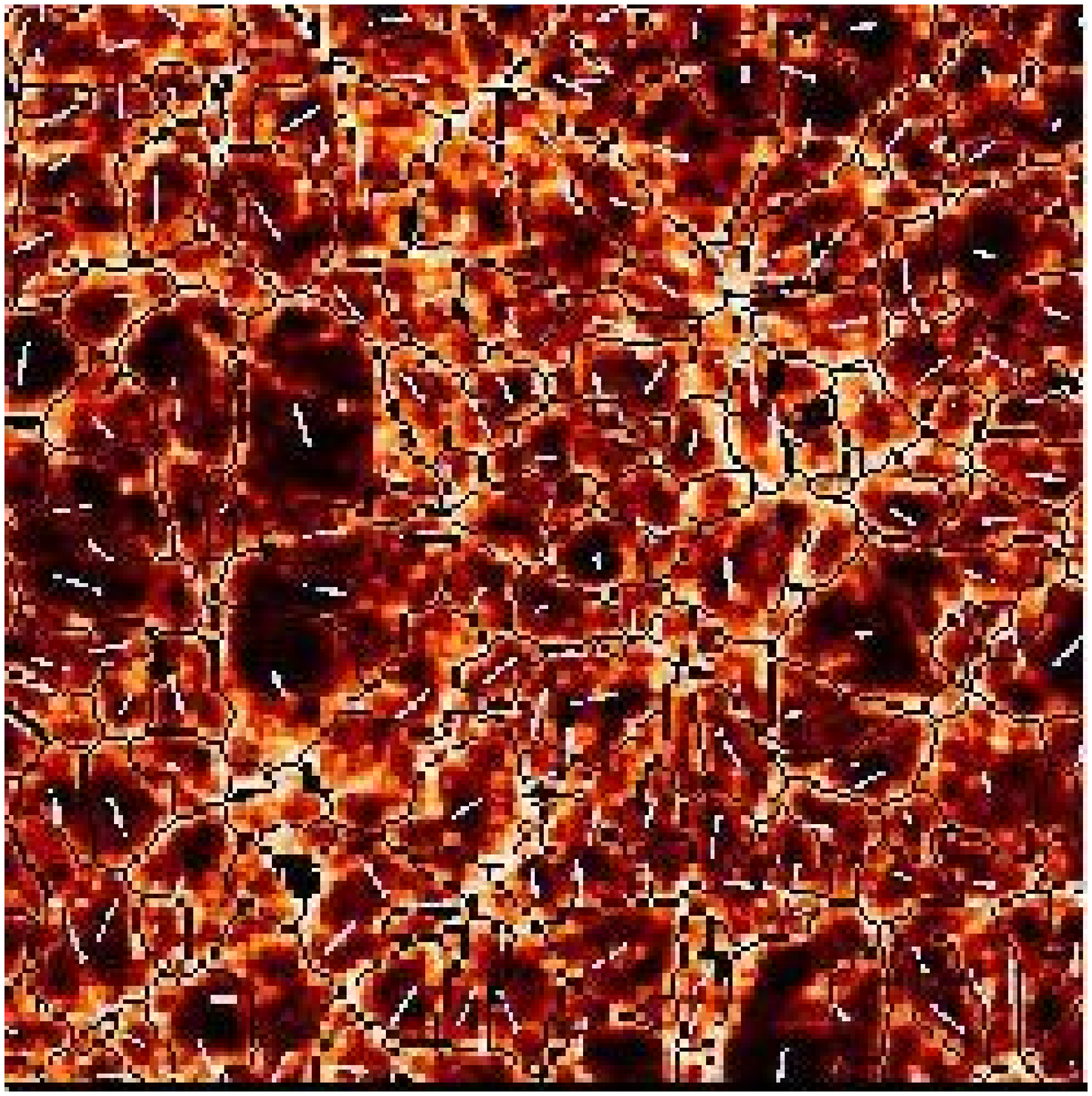}
  \vspace{0.5cm}\\
  \includegraphics[width=0.45\textwidth]{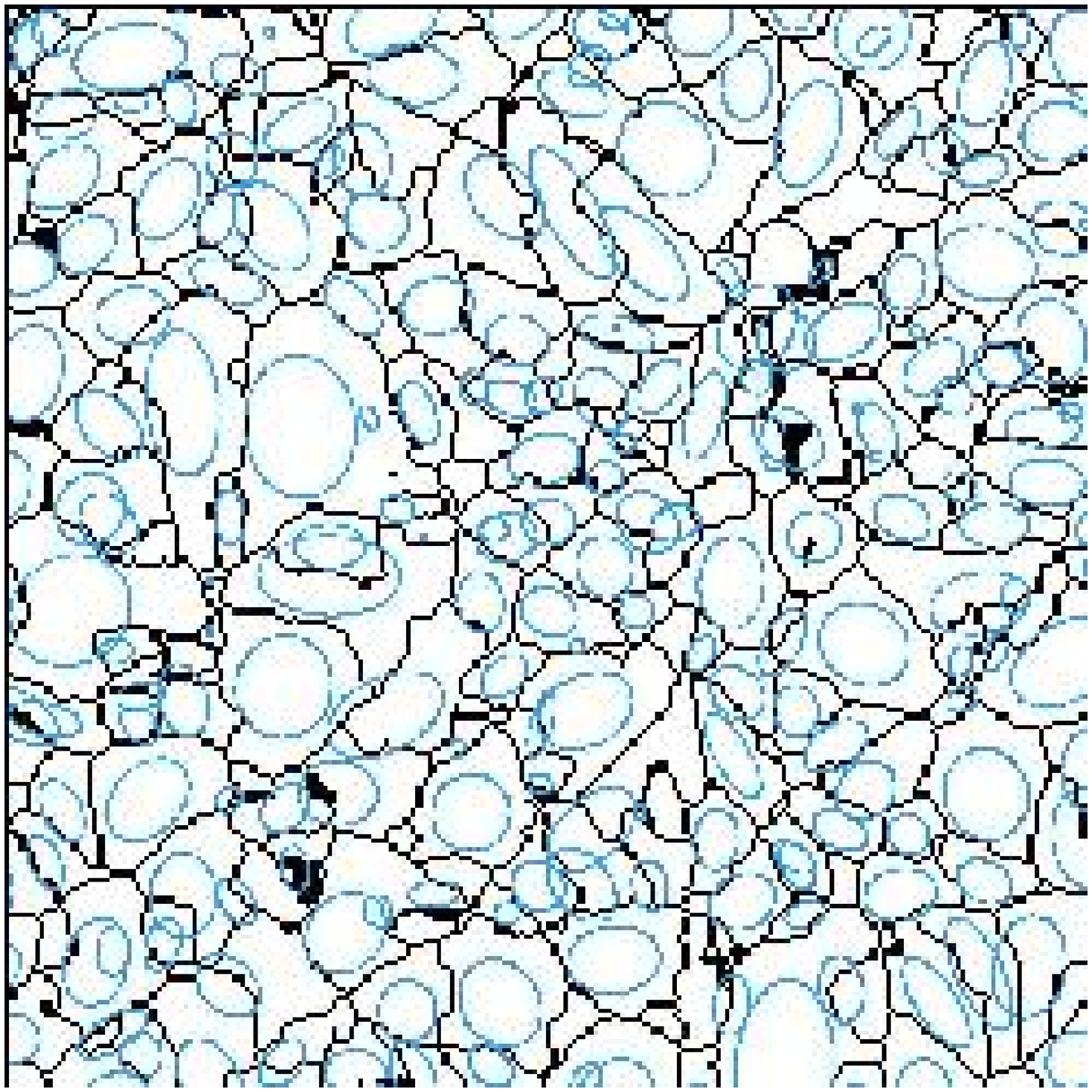}\hspace{0.5cm}
  \includegraphics[width=0.45\textwidth]{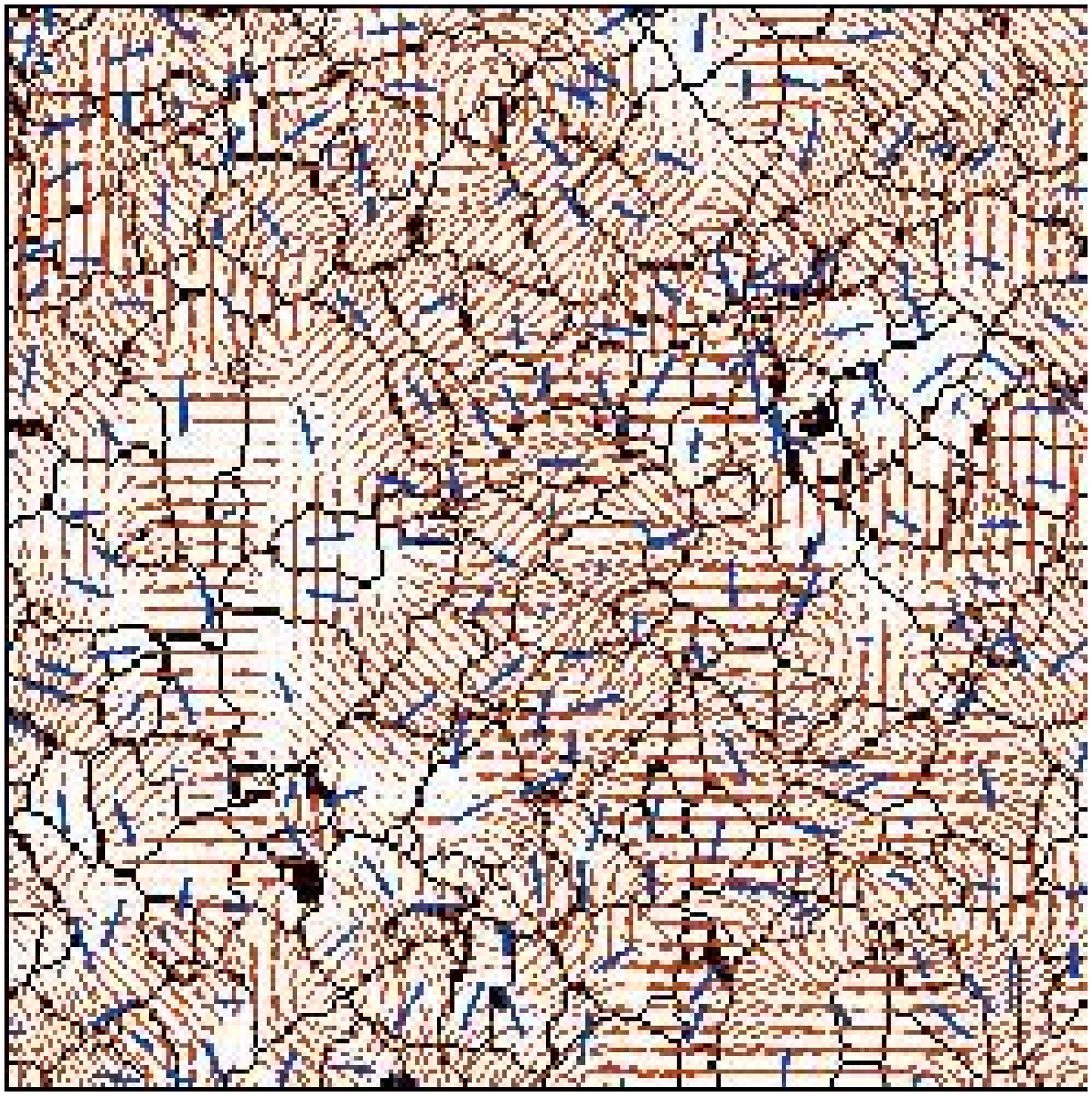}
 
  \caption{The four panels show a slice through the density field with the void-segmentation superposed. 
  The first panel (a) a slice through the entire simulation. The image shows the density field of 
  the simulation. The density field is determined by DTFE. Superimposed are the black solid lines 
  indicating the boundaries of the WVF voids. The white box indicates the section which forms the 
  subject of the next three frames. Second panel: density field in the subbox with the WVF void 
  boundaries superimposed (solid lines). The blue bars indicate the direction of the projection 
  of the longest void axis. Third panel: the WVF void boundary contours (solid lines) with the 
  fitted void shape ellipsoids sperimposed. Fourth panel: on the landscape with WVF void boundaries 
  the tidal field compressional component is represented by tidal bars (red), representing the 
  direction and strength of the tidal field. Also depicted are the void shape bars (blue). }
  \label{fig:slices}
\end{figure*}
 
Voids are a dominant component of the Cosmic Web \citep[see
e.g.][]{tully2007, weyrom2007}, occupying most of the volume of space.
In this paper we present evidence for significant alignments between
neigbouring voids and establish the intimate dynamic link between
voids and the cosmic tidal force field.  The Watershed Void Finder
(WVF) procedure \citep{platen2007} that we use to identify voids is
crucial to the succes of this analysis.  The WVF technique is based on
the topological characteristics of the spatial density field and
thereby provides objectively defined measures for the size, shape and
orientation of void patches.
 
Large voids form around deep density troughs in the primordial density
field.  The main aspects of the evolution of a large void may be
understood on the basis of the expansion of simple spherical and
isolated under-densities (e.g. \cite{edbert1985}).  Under-dense
regions expand with respect to the background Universe and in general
will have the tendency to grow more spherical with time
\citep{icke1984}.  However, \cite{shandarin2006} and
\cite{parklee2007a} have emphasized that, in realistic cosmological
circumstances, voids will be nonspherical.  This is quite apparent in
images of, for example, the Millennium simulation
\citep{springmillen2005}.  Moreover, substructures within voids
display a manifest alignment along a preferred direction.  This is, in
part, a consequence of the relatively strong influence of the
surrounding inhomogeneous matter distribution on the void's structure
and evolution.
 
Voids evolve hierarchically: as they expand with respect to the
background Universe they merge with their surrounding peers, building
voids of ever larger sizes \citep{regoes1991, weykamp1993,
gottloeber2003, colberg2005}.  Small voids can survive as substructure
within the larger voids or may disappear under the gravitational
impact of surrounding overdense structures \citep{sahni1994,
shethwey2004}.  The result is a void population whose scale is
distributed around a characteristic void size: we shall show this
using our WVF sample.
 
\begin{figure*}
     \mbox{\hskip -0.6truecm\includegraphics[width=0.36\textwidth]{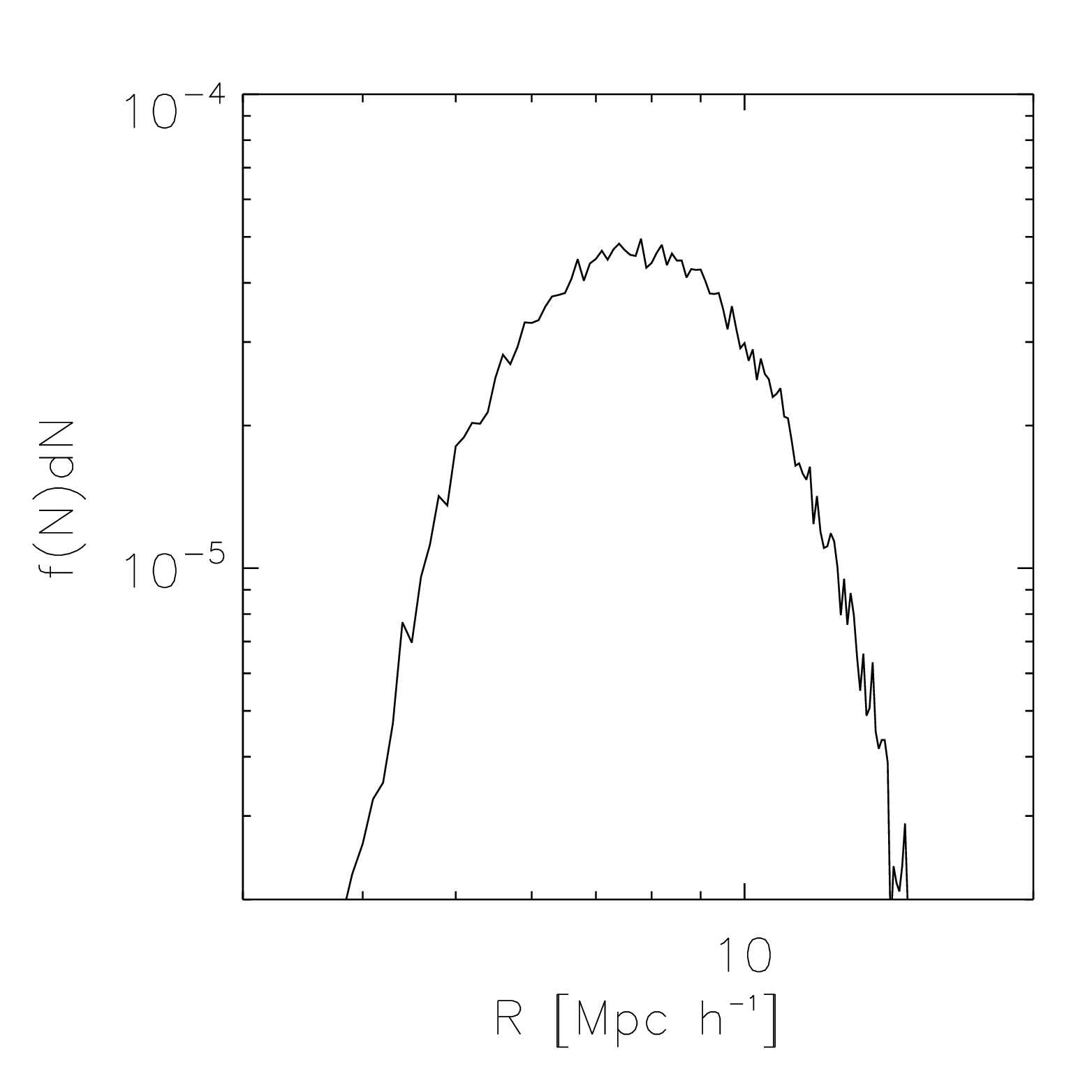}}
     \mbox{\hskip -0.5truecm\includegraphics[width=0.36\textwidth]{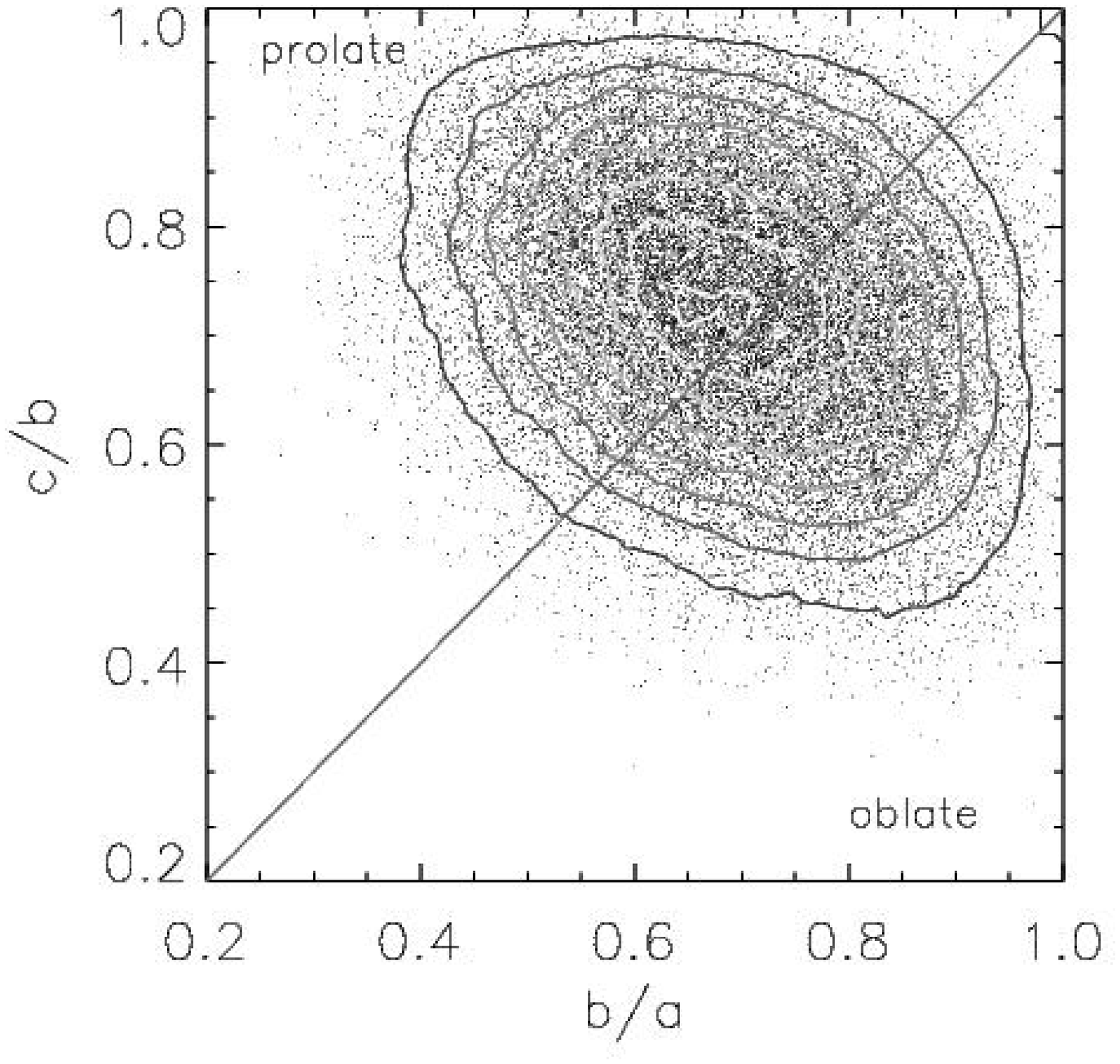}}
     \mbox{\hskip -0.5truecm\includegraphics[width=0.36\textwidth]{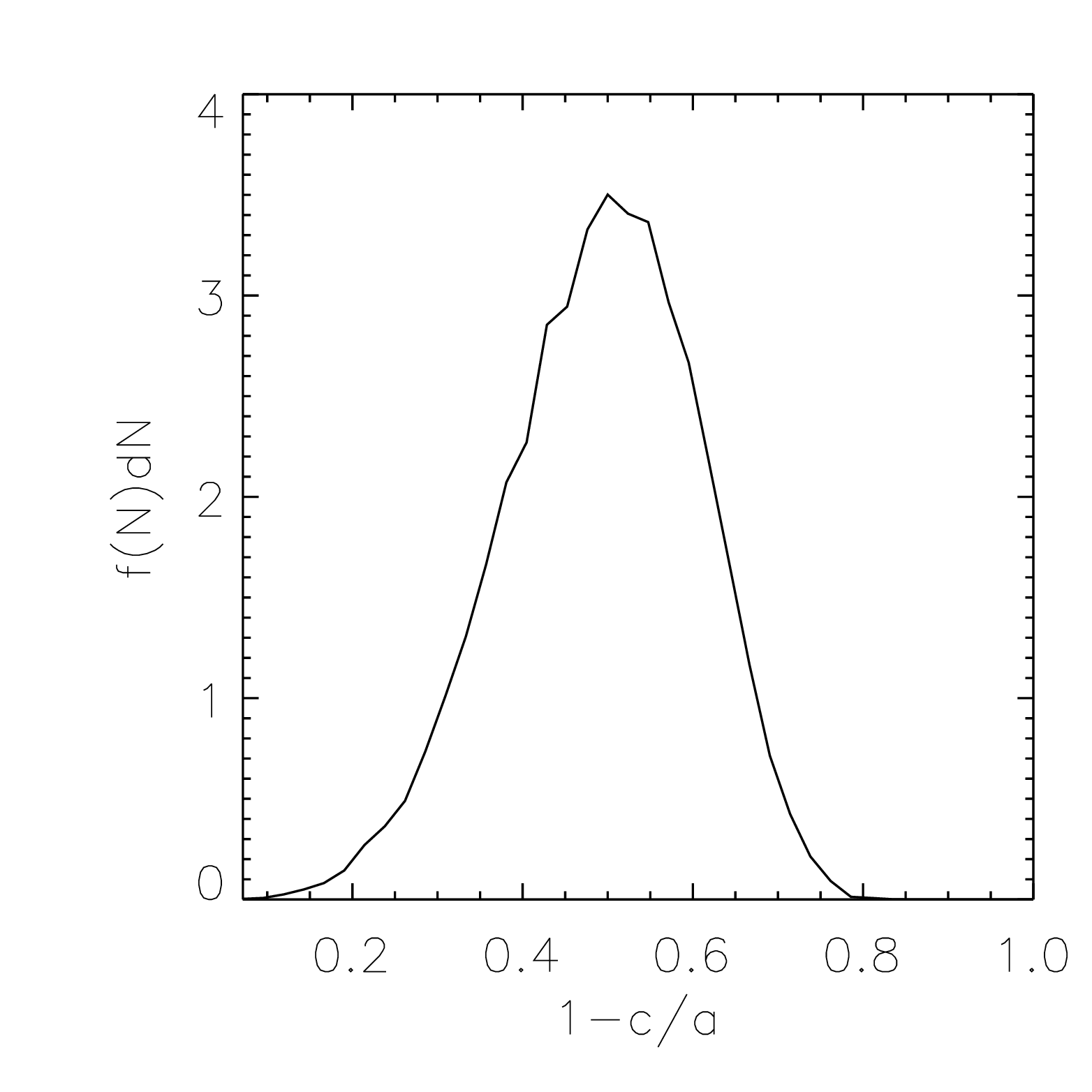}}
     \caption{Void characteristics and statistics. Left: the (peaked) void size distribution as a function of 
      void radius $R_v$. Centre: scatter diagram of the two void axis ratios $\eta_{32}=a_3/a_2$ versus 
      $\eta_{21}=a_2/a_1$. Right: distribution of void ellipticity $\epsilon=1-a_3/a_1$.}
     \label{fig:shape}
\end{figure*}
A major manifestation of large scale tidal influences is that of the
alignment of shape and angular momentum of objects
\citep[see][]{bondweb1996, desjacques2007}.  The alignment of the
orientations of galaxy haloes, galaxy spins and clusters with larger
scale structures such as clusters, filaments and superclusters have
been the subject of numerous studies \citep[see e.g.][]{binggeli1982,
bond1987, rhee1991, plionis2002, basilakos2006, trujillo2006,
aragon2007, leevrard2007, parklee2007b,leespringel2007}.  Recent
analytical and numerical work by \cite{parklee2007a} discussed the
magnitude of the tidal contribution to the shape of voids \citep[see
also][]{leepark2006}.  Describing the evolution of voids under the
influence of tidal fields by means of the Lagrangian Zel'dovich
approximation, they found that the ellipticity distribution of voids
is a sensitive function of various cosmological parameters and
remarked that the shape evolution of voids provides a remarkably
robust constraint on the dark energy equation of state
\citep{leepark2007}.
 
In section 2 we introduce the N-body simulation which we used to
identify the voids, along with a brief description of the WVF method
we use to identify the voids in our sample. The results for individual
void characteristics, in particularly the size and shape of voids, are
presented in section 3.  Then in section 4 we present the analysis of
the alignment of voids in the sample.  The possible origins for the
void alignment are discussed in section 5, and in section 6 we focus
on the role of tidal fields in determining the evolutionary history of
voids.  Finally, section 7 concludes with a short discussion and
prospects.
 
\section{The Void Sample}
It is essential to have a statistically representative void sample for
our study of the shapes of voids and their relative alignment.  We
also need a sufficiently large spatial coverage in order to evaluate
tidal field contributions over a large range of scales.  Thus it is
necessary to have a simulation with a large box size.  We used the VLS
simulation from the Virgo Supercomputing Consortium\footnote{The Virgo
Consortium VLS $\Lambda$CDM data is available at
http://www.mpa-garching.mpg.de/galform/virgo/vls/}, a simulation of
cosmic structure formation in a $\Lambda$CDM cosmology with
$\Omega_{m,0}=0.3$, $\Omega_{\Lambda,0}=0.7$, $H_0=70 \: \mathrm{km \:
s}^{-1} \mathrm{Mpc}^{-1}$ and $\sigma_8=0.9$.  The simulation
contains $512^3$ N-body particles, each with a mass of $6.86 \times
10^{10}h^{-1} M_{\odot}$ in a comparatively large box of side $479
h^{-1} \mathrm{Mpc}$.  We used the particle distribution at the final
timestep, corresponding to the current cosmic epoch $a=1$.
 
\subsection{Void Identification}
The particle distribution of the simulation is converted into a
grid-based density field by means of Delaunay Triangle Field Estimator
(DTFE) \citep{weyschaap2007}.  The key feature of DTFE is that it is a
reconstruction method that does not involve any user-defined
parameters.  Moreover, its natural and intrinsic filtering and
interpolation procedure yields a density field that remains faithful
to both the density and shape structure of the original particle
distribution. DTFE guarantees an optimal representation of the
geometry and topology of the weblike patterns in the mass
distribution. A slice through the DTFE resampled density field is
shown in the first panel of figure~\ref{fig:slices}.
 
After resampling the point density distribution with DTFE, we applied
the Watershed Void Finder (WVF) \citep{platen2007} in order to
identify the voids in the simulation.  The WVF procedure traces the
basin and its encompassing boundary for each minimum in the spatial
mass distribution.  Because WVF is free of any a priori shape
criterion it is ideally suited to analyzing the morphology of
underdense regions in the cosmic web.
 
The WVF procedure works as follows \citep[see][for a detailed
description]{platen2007}. First we remove shot noise effects in the
raw DTFE density field by smoothing at the Nyquist scale ($\approx 1.0
h^{-1}\mathrm{Mpc}$); this is done using a Butterworth $n=2$ filter.
Then we identify the minima in the resulting density field and tag
each minimum with a unique identification number.  We can visualize
the density field as a landscape with basins around each minimum.
Starting from the minima we flood this landscape.  As the flood level
rises the basins surrounding each miminum start to fill up.  Walls
gradually emerge along the crest lines of the density field at the
interstices between distinct regions. The final product of the
procedure will be a division of the density field into individual void
cells.
 
\section{Void Characteristics}
In the first panel of figure~\ref{fig:slices} each void identified by
WVF is indicated by the solid black line marking the surrounding void
boundary.  The image shows that we indeed have a large number of voids
in our sample. In the upper right-hand panel we zoom in on a region of
the simulation box (the white box in first panel) which we use for the
purpose of illustrating the subsequent steps (the analysis is done on
the entire box).  The upper right panel also has lines indicating the
(projected) magnitude and direction of the major axis of each void.
Visual inspection of the void regions in this panel manifestly
suggests some degree of alignment. Quantitative analysis will
establish the significance of that alignment.
 
\subsection{Void Size}
The size and shape distribution of the WVF-identified voids is shown
in figure~\ref{fig:shape}. In the simulation we found 29000 voids,
with an average void size $R_V \approx 8.5 h^{-1} \mathrm{Mpc}$.  The
void size distribution in figure~\ref{fig:shape}a clearly shows a peak
in the size distribution, confirming the prediction by
\cite{shethwey2004}.
 
\subsection{Void Shape}
For each individual void region we calculate the shape-tensor
$\mathcal{S}_{ij}$ by summing over the $N$ volume elements $k$ located
within the void,
\begin{eqnarray}
  \qquad \mathcal{S}_{ij}& = - &\sum_{k} x_{ki} x_{kj} \qquad \ \ \ \ \ \ \ \ \textrm{(offdiagonal)} \\
  \qquad \mathcal{S}_{ii}& =   &\sum_{k} \left({\bf x}_k^2-x_{ki}^2\right) \qquad\ \ \textrm{(diagonal)}\,,\nonumber 
\end{eqnarray}
where ${\bf x}_k$ is the position of the $k$-th volume element within the void, with respect to the (volume-weighted)  
void center 
${\bf {\overline{r}}}_v$, i.e. 
${\bf x}_k = {\bf r}_k - {\bf {\overline{r}}}_v$.  
The shape tensor $\mathcal{S}_{ij}$ is related to the inertia tensor $\mathcal{I}_{ij}$.  However, it differs in assigning equal weight to each volume element within the void region. Instead of biasing the measure towards the mass concentrations near the edge of voids, the shape tensor $\mathcal{S}_{ij}$ yields a truer reflection of the void's interior shape. 
 
When evaluating alignments we will mostly use the traceless version of the shape tensor:
\begin{equation}
{\tilde {\mathcal{S}}}_{ij} = {\mathcal{S}}_{ij}\,-\,\frac{1}{3}\,{\mathcal{S}}\,\delta_{ij}\,,\qquad
\mathcal{S} = \sum_{k} {\mathcal{S}}_{kk}
\end{equation}
The length and orientation of the shape ellipsoid are inferred from the eigenvalues and eigenvectors of $\mathcal{S}_{ij}$.  The eigenvalues of the void ellipsoid are labelled so that $s_1 > s_2 > s_3$. The corresponding semi-axes are oriented along the directions of the corresponding unit eigenvectors ${\hat {\bf e}}_{s1}$, ${\hat {\bf e}}_{s2}$ and ${\hat {\bf e}}_{s3}$. The length of the semi-axes $a_i$ are 
\begin{equation}
a_1^2\,=\,{\displaystyle 5 \over \displaystyle 2N}\,\left\{s_2\,+\,s_3\,-\,s_1\right\}
\end{equation}
with cyclic permutation for the other axes ($N$ is the number of volume elements within the void). The longest axis is directed 
along ${\hat {\bf e}}_{s1}$, the shortest along ${\hat {\bf e}}_{s3}$. 
 
We quantify the shape of the void ellipsoids in terms of the two axis ratios $\eta_{21} = a_2/a_1$ and $\eta_{32} = a_3/a_2$.  Amongst the triaxial void ellpsoids we make a distinction between oblate ones, with $\eta_{21}>\eta_{32}$, and prolate ones having $\eta_{21}<\eta_{32}$.  If voids were spherical we would have $\eta_{21}=\eta_{32}=1$ (though, in reality, perfect sphericity does not exist).
 
\subsection{Results on Void Shapes}
In the density field in the upper righthand frame of
figure~\ref{fig:slices} we have indicated the projected direction of
the longest axis of the voids by means of a bar along the
corresponding eigenvector, with the length of the bar proportional to
the size of the void axis. It allows a superficial comparison with the
related WVF void patches (indicated by the solid black lines marking
their boundary).
 
The lower lefthand panel of figure~\ref{fig:slices} shows the
shape-ellipsoids of each void within the region of the simulation box
shown in the top-right panel.  The centers of the shape-ellipsoids are
located at the center of each void with their size scaled according to
match the void volume: they give a convenient visual impression of the
void shapes and sizes.  The reasonably accurate degree to which the
ellipsoids reflect the shape of the voids may be inferred from a
comparison with the WVF void regions themselves.
 
Clearly, the voids are quite nonspherical. A more quantitative
impression of the shape distribution may be found in the second and
third panel of fig.~\ref{fig:shape} which show that the voids are far
from spherical.
 
The intrinsic triaxial shape of voids is apparent from the
distribution in the second panel of fig.~\ref{fig:shape}. This shows a
scatter plot of of the two axis ratios, $\eta_{12}$ versus $\eta_{23}$
for every void in the sample.  To guide the eye we have superimposed
the inferred isodensity contours defined by the point density in the
scatter plot.  We find a slight asymmetry: there are more prolate
voids than oblate ones.
 
The third panel shows that the distribution of the ellipticity
$\epsilon = 1-a_3/a_1$ is skewed towards higher values of $\epsilon >
0.5$.  The void population is marked by pronounced triaxial shapes:
the average ratio between the smallest and largest axis $c/a \approx
0.49$.  This ratio agrees with well with the value of $0.45$ found by
\cite{shandarin2006} (which is perhaps a little surprising given the
totally different void definitions used in that paper).
 
In summary, we find voids to be slightly prolate, with average axis ratios in the order of $c:b:a\approx 0.5:0.7:1$. 
 
Two important factors contribute to this flattening.  Even though,
internally, voids tend to become more spherical as they expand
\citep{icke1984}, perfect sphericity will hardly ever be reached.
Before voids would be able to become spherical they would encounter
surrounding structures such as overdense filaments or planar walls.
Even more important may be the fact that, for voids, external tidal
influences are particularly important: voids will always be rather
moderate underdensities since they can never be more underdense than
$\delta=-1$.  The external tidal forces drive a significanr anisotropy
in the development of the voids, and in the extreme cases may cause
complete collapse of the void along one or more of its axes.  This is
another aspect of the way in which tidal forces shape the Cosmic Web,
as has been emphasized by \cite{bondweb1996}.
 
\section{Void Alignments}
Large scale influences play a major role not only in shaping
individual voids but also on their mutual arrangment and
organization. A particularly important manifestation of this is the
mutual alignment between voids.  In this section we establish the
significance of this effect, and in the next section we analyse the
correlation between the tidal force field and the void orientation.
 
\subsection{Void Alignments: definitions}
We take the void centers in our sample to define a spatial point
process.  Three void alignment measures have been investigated. Each
is a marked correlation function \citep{beiker2000, stoystoy1994} and
assesses the degree of alignment as a function of the distance $r$
between the void centres. The mark for each void pair is the cosine of
the angles between two specific axes of the void or a combination of
these.
 
The first alignment measure is the full shape correlation ${\mathcal
A}_{SS}$.  The second measure concerns the alignment strength
${\mathcal A}_{33}$ between the longest axes of the voids and the
third measure concerns the alignment strength $A_{11}$ between the
shortest axes of the voids. To determine these functions we evaluate
the alignment between each pair of voids as a function of the
separatin of their centers $r$.
 
The alignment between a pair of voids $m$ and $n$, with traceless
shape tensors ${\tilde {\mathcal{S}}}_{m,ij}$ and ${\tilde
{\mathcal{S}}}_{n,ij}$, is specified in terms of the following set of
quantities,
\begin{eqnarray}
{\Gamma}_{SS}(m,n)&\,=\,&{\displaystyle \sum\nolimits_{i,j} 
{\tilde {\mathcal S}}_{m,ij}\,{\tilde {\mathcal S}}_{n,ij} \over 
\displaystyle {\tilde {\mathcal S}}_m\,{\tilde {\mathcal S}}_n}\nonumber\\
\ \\
\Gamma_{11}(m,n)&\,=\,&{\hat {\bf e}}_{s1}(m)\cdot {\hat {\bf e}_{s1}}(n)\,\nonumber\\
\Gamma_{22}(m,n)&\,=\,&{\hat {\bf e}}_{s2}(m)\cdot {\hat {\bf e}_{s2}}(n)\,\nonumber\\
\Gamma_{33}(m,n)&\,=\,&{\hat {\bf e}}_{s3}(m)\cdot {\hat {\bf e}_{s3}}(n)\,,\nonumber
\end{eqnarray}
in which ${\tilde {\mathcal S}}_m$ and ${\tilde {\mathcal S}}_n$ are
the norms of the corresponding (traceless) shape-ellipsoids. In
essence $\Gamma_{11}$, $\Gamma_{22}$ and $\Gamma_{33}$ are the cosines
of the angles between the corresponding void axes, directed along the
shape eigenvectors ${\hat {\bf e}}_{s1}$, ${\hat {\bf e}}_{s2}$ and
${\hat {\bf e}}_{s3}$.  These alignment measures do not depend on the
degree of ellipticity of each void, they focus exclusively on the
relative orientation of the void axes.
 
The corresponding marked correlation functions are an expression of the strength of the alignments as a function of distance $r$ between the void centers, 
\begin{eqnarray}
&{\mathcal C}_{SS}(r) = \langle\,\Gamma_{SS}\,\rangle \\
&{\mathcal C}_{11}(r) = \langle\,\Gamma_{11}\,\rangle\, ,\;
{\mathcal C}_{22}(r) = \langle\,\Gamma_{22}\,\rangle\, ,\;
{\mathcal C}_{33}(r) = \langle\,\Gamma_{33}\,\rangle\nonumber
\label{eq:voidcorr}
\end{eqnarray}
The brackets indicate the ensemble average over all void pairs $(m,n)$ whose mutual separation $r$ lies within a band $\Delta r$ around $r$, $|{\bf {\overline{r}}}_m-{\bf {\overline{r}}}_n| \in [r,r+\Delta r]$.  In practice, we bin the void pairs in our void sample into bins which each contain 5000 pairs.  This ensures that each bin has equal statistical significance. 
 
Note that because the angle between the corresponding void ellipsoid
axes of two voids $m$ and $n$ has a value in the range $0 \leq
\theta(m,n)\leq \pi/2$, the alignment $\Gamma_{ii}(m,n)$ between these
two voids lies within the range $0\leq \Gamma_{ii}(m,n)=\cos
\theta_{mn} \leq 1$.  Each of the axis correlation functions
${\mathcal C}_{ii}$ (i=1,2,3) is the average cosine between
corresponding two void axes, ${\mathcal C}_{ii}=\langle \cos \theta
\rangle$. For a perfectly randomly oriented sample we would therefore
expect ${\mathcal C}_{11}={\mathcal C}_{22}={\mathcal C}_{33}=0.5$.
In the case of perfect alignment we would obtain ${\mathcal
C}_{ii}=1.0$.  A value of $C_{ii}\approx 0$ would suggest an
anti-alignment, a cross pattern, corresponding to an orientation angle
$\theta_{mn}\approx \pi/2$.
 
\subsection{Void Alignments: significance}
We can gain insight into the significance of the values of the
alignment measures ${\mathcal C}_{ii}$ by comparing the values
obtained for a sample of vectors in which a perfectly aligned subset
is mixed with a entirely randomly oriented set of voids.
 
In this comparison sample, suppose a fraction $f_r$ of the void sample
has a random orientation while the remaining fraction $f_a=1-f_r$ has
a singular orientation along one direction. We may then infer the
implied random fraction $f_r$ in the comparison sample for the
obtained values for ${\mathcal C}_{ii}$ in our real void sample.
 
It is straightforward to infer the implied values of the alignment measures ${\mathcal C}_{ii}$ in the comparison sample: 
\begin{eqnarray}
{\mathcal C}_{ii}\,=\,\langle \cos\theta \rangle&\,=\,&0.5 f_r^2\,+\,f_r f_a\,+\,f_a^2 \nonumber \\ 
&\,=\,&0.5\,(f_r-1)^2\,+\,0.5.  
\end{eqnarray}
For example, if we were to have $20\%$ aligned voids mixed with $80\%$
random ones, the alignment would be ${\mathcal C}_{ii}=0.52$. An
alignment of ${\mathcal C}_{ii}=0.63$ would correspond to $\approx
50\%$ aligned and $\approx 50\%$ random voids, while ${\mathcal
C}_{ii}=0.75$ would imply $\approx 70\%$ aligned and $\approx 30\%$
random ones.
 
\subsection{Void Alignments: results}
\begin{figure}
     \mbox{\hskip -0.7truecm\includegraphics[width=0.53\textwidth]{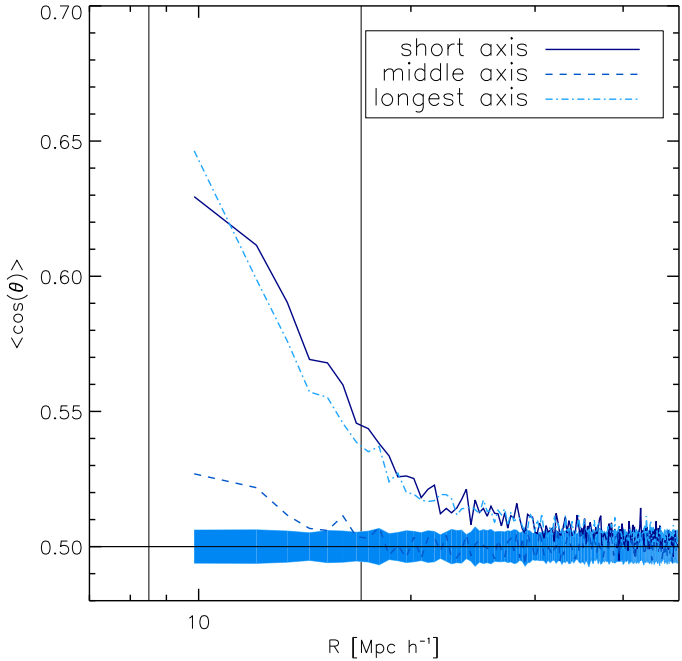}}
     \caption{The alignment of the three void axes. The figure shows the (marked) correlation 
       functions ${\mathcal C}_{11}(r)$ (solid), ${\mathcal C}_{22}(r)$ (dashed) and 
       ${\mathcal C}_{33}(r)$ (dot-dashed), the ensemble average of the cosine of the angle between 
       the longest, medium and shortest axes of voids at a distance $r$. A pure random orientation 
       corresponds to ${\mathcal C}_{ii}(r)=0$, a perfectly aligned one ${\mathcal C}_{ii}(r)=1$. 
       The coloured bar around the value ${\mathcal C}_{ii}=0.50$ indicates the $2-\sigma$ 
       dispersion of a similar random sample. }
     \label{fig:cor1}
\end{figure}

Figure~\ref{fig:cor1} presents the results for the measured
correlations of the void orientations.  Each curve corresponds to the
orientation correlation between one of the three void axes of voids:
the solid black curve to that of the longest axis, ${\mathcal
C}_{33}$, the dot-dashed curve to that of the shortest axis,
${\mathcal C}_{11}$, and the intermediate axis ${\mathcal C}_{22}$
(dashed curve).
 
The figure shows the longest and the smallest axis exhibit the
strongest alignment. At short distances their alignment is very strong
and remains substantial out to relatively large radii of $20 - 30
h^{-1}\hbox{\rm Mpc}$. Beyond this distance the alignment rapidly
declines towards random values. At the smallest distances probed the
alignment of the longest and shortest axis reaches values of around
0.65. This corresponds to a random component of only 30 percent, in
combination with 70 percent perfectly aligned voids. By contrast, the
second axis shows a striking lack of alignment on all scales and,
except at the shortest separations, appears to be mostly randomly
oriented. This is partially due to the considerable scatter in the
orientation of the intermediate axis. This may be a reflection of the
crudeness of approximating the void shapes by ellipsoids. However,
even taking account of the implied scatter the intermediate axis
remains more weakly correlated.
 
To provide an estimate of the significance of the results we have
estimated the standard deviation expected for a sample of perfect
randomly oriented voids.  The $2 \sigma$ spread is depicted by means
of the coloured bar around the expectation value for a random sample,
$C_{ii}=0.5$. We obtained these estimates by a Monte Carlo experiment
in which we randomly shuffled the shape ellipsoids over the the voids
and subsequently re-measured the alignment. Note that the variance
remains nearly constant over the whole range. This is a consequence of
our decision to define bins having equal number of particles per bin.
 
\section{The cause of Void Alignment}
The possibility that the alignment may be an artefact of the WVF
method has to be considered: large spherical voids that were chopped
in half would provide an anisotropic void set in which neighbouring
voids were strongly aligned. However, this it can quickly be
eliminated simply by examining figure~\ref{fig:slices} which shows no
such phenomenon.  We shall later on present further evidence based on
tidal fields that WVF is not itself the source of the alignment.

The reason for the alignment between voids may be due to any of a
number of physical effects. Here, we consider three possible physical
sources for this alignment: 1) the initial conditions, 2) the
geometric packing of the voids and 3) the large scale tidal field.
 
\subsection{Initial conditions} 
The first possibility, that the structural correlations are born with
the primordial density field, is difficult to address since, at least
in the linear regime, today's matter distribution a direct reflection
of those initial conditions.  Thus the density fluctuations and the
tidal fields that distort them are both present initially, either
being deducible from the other.  The Cosmic Web, voids, galaxy halos
as well as the tidal field are all manifestations of the same initial
matter distribution.
 
Voids emerge out of a primordial Gaussian density field. Density peaks
and dips in a Gaussian random field are not only clustered
\citep{bbks}, they are also mutually aligned \citep{bond1987}. This
alignment stretches out over scales over which they possess a nonzero
correlation function, approximately the same range as that of the
comparable population of rich clusters. Even when their evolution
would not be influenced by external forces the voids would retain the
memory of their initial configuration.  Their initial shape would be
(slightly) attenuated by their internally driven expansion and the
void would keep its initial orientation.
 
The primordial alignment of density peaks is enhanced by subsequent
nonlinear interactions with surrounding matter condensations. Voids,
however, will be far less affected due their relatively large spatial
extent and relaively low density contrast. If anything, their initial
alignment will be strengthened by the influence of the large scale
tidal field.
 
\subsection{Void packing}
Voids are a unique constituent of the cosmic inventory in that they
represent a volume-filling population
\citep{shethwey2004}. Consequently, the placement of voids in the
large scale matter distribution is a nontrivial issue related to the
random object packing problem in mathematics and material's science
\citep[see e.g.][]{conwaysloane1998}.
 
We have argued before that an important consequence of the tight
packing of voids concerns their shape. As they try to expand outward
they will meet up with the surrounding large scale structures. This
involves the surrounding filaments, walls and clusters as well as
their neigbouring voids. This strongly restricts their tendency to
become more spherical as they expand \citep{icke1984, edbert1985}.
While the inner cores of voids \citep{weykamp1993} remain spherical,
the surroundings mould them into a generally more flattened
configurations.  (The shape of regions defined by various level-sets
in the initial density field is discussed in detail by
\cite{shandarin2006}).
 
Mathematical studies of the packing problem focus mainly on the
packing of equivalent spheres. The more complex problem of ellipsoid
packings was recently adressed in a seminal experimental study on the
tight packing of (oblately shaped) M\&M's \citep{donev2004}. Not only
did they prove that such ellipsoids allow a tighter packing than
feasible with purely spherical objects, but also that this goes along
with a configuration of strongly aligned ellipsoids \citep[see
fig. 1B,][]{donev2004}.
 
The situation for the packing of voids is considerably more complex.
Cosmic voids have, as we have shown, a considerable diversity in size
and shape.  Moreover, unlike M\&M's, voids are not solid bodies: the
void ellipsoids may not be entirely contained within the boundaries of
their void patches and as a result may overlap.  Nonetheless, we might
on purely geometric grounds expect an alignment of a void with its
``natural'' neighbours and that, locally, the orientation of voids is
strongly constrained by the shape and orientation of its neighbours.
 
In order to further assess the situation we have analyzed four
different configurations: a $\Lambda$CDM initial Gaussian field, a
Poisson field, a regular crystalline configuration and an irregular
Voronoi cellular configuration.  These reflect one or more aspects of
the genuine void distribution (an accompanying publication will
provide the details of the study).
 
In each of these special configurations we see a rapid decline in
alignment strength beyond the characteristic void size $R_V$, even
turning into anticorrelation around $2R_V$.
 
Figure~\ref{fig:cor1} shows the large scale correlations for void
alignment in the N-Body model we have studied.  We have marked the
scales $R_V$ and $2R_V$ by means of two vertical lines.  The fact that
in the void sample of this study we find significant alignments beyond
$2R_V$ is a strong argument for the alignments being due to other
factors.
 
\subsection{Large Scale Tidal Forces}
The intricate patterns that constitute the cosmic web are a direct
product of the large scale tidal field \citep{bondweb1996}. Voids are
an integral part of this pattern and their shape and orientation is
expected to react strongly to the surrounding inhomogeneous matter
distribution via their tidal influence. The tidal field is a
reflection of the underlying inhomogeneous cosmic matter distribution
and contains contributions from matter inhomogeneities over a large
range of scales.
 
In a sense, the influence of the tidal field on void orientations is
strongly coupled to the initial conditions (see above): already in the
initial density field the local shape of a density trough is
considerably correlated with the tidal field configurations over a
large range of scales.
 
The key issue is the influence of tidal forces on the evolution of the
shape and orientation of voids. This includes the impact on the
nonlinear development of the void. We are particularly interested in
identifying the largest scales over which we can recognize a clear
alignment between the tidal field and the orientation of the
voids. Their spatial coherence will be crucial for the void-void
alignments.
 
In the next section we will focus on this specific issue. 
 
\section{Influence of Tidal Field}
To test the relationship between the large scale tidal force field and
the void correlations of which they are supposed to be the source we
investigate the correlation, at the location of each of the voids,
between the tidal tensor ${\mathcal T}_{ij}$ and the shape tensor
${\mathcal S}_{ij}$.
 
\subsection{Tidal Field}
We compute the traceless tidal shear components,
\begin{equation}
  \mathcal{T}_{ij}\,=\,{\frac{\partial^2 \phi}{\partial r_i\partial r_j}\,-\,{\frac{1}{3}}\,\nabla^2\phi\,\delta_{ij}}\,.
\end{equation}
for the entire simulation volume.  The gravitational potential $\phi$,
due to the underlying density distribution, is determined from the
DTFE density field $\delta({\bf r})$ by solving the Poisson equation
in Fourier space.
 
The eigenvalues of the tidal tensor are ordered as $t_1 > t_2 > t_3$,
with corresponding unit eigenvectors ${\hat {\bf e}}_{t1}$, ${\hat
{\bf e}}_{t2}$ and ${\hat {\bf e}}_{t3}$.  We also draw a distinction
bewteen the compressional component of the tidal field, that part of
the tidal field corresponding to the positive eigenvalues of
${\mathcal T}_{ij}$, and the dilational component. The compressional
component always includes the component $t_1{\hat {\bf e}}_{t1}$ while
the dilational component always includes $t_3{\hat {\bf
e}}_{t3}$. Depending on its sign, $t_2 {\hat {\bf e}}_{t2}$ may
contribute to either the compression or to the stretching of the mass
element.
 
\subsubsection{Tidal-Tidal alignments}
In our bid to identify the source of the void-void shape correlations
we also need to understand the auto-correlations between the tidal
field contributions originating at different scales. To this end we
define a {\it tide-tide} alignment measure ${\mathcal
A}_{TT}(R_1,R_2)$.
 
At a location ${\bf r}$ in the density field we may compare the orientation of the tidal field filtered at two different scales $R_1$ and $R_2$, yielding the alignment
\begin{equation}
{\Gamma}_{TT}({\bf r};R_1,R_2)\,=\,\ {\displaystyle \sum\nolimits_{i,j} 
{\mathcal T}_{ij}({\bf r};R_1)\,{\mathcal T}_{ij}({\bf r},R_2) 
\over \displaystyle {\mathcal T}({\bf r};R_1)\,\,{\mathcal T}({\bf r};R_2)}
\end{equation}
\noindent where ${\mathcal T}({\bf r};R_1)$ is the norm of the tidal tensor ${\mathcal T}_{ij}({\bf r};R_1)$ filtered on a scale $R_1$ and ${\mathcal T}({\bf r};R_2)$ on scale $R_2$. 
 
The ensemble average of ${\Gamma}_{TT}({\rm r};R_1,R_2)$ is the tide-tide alignment ${\mathcal A}_TT(R_1,R_2)$,
\begin{equation}
{\mathcal A}_{TT}(R_1,R_2)\,=\,\langle\,{\Gamma}_{TT}(R_1,R_2)\,\rangle
\label{eq:att}
\end{equation}
This quantity is computed by taking the average of ${\Gamma}_{TT}({\bf r};R_1,R_2)$ over a large number of locations ${\bf r}$.
 
\subsubsection{Void-Tidal Field alignment: formalism}
In order to trace the contributions of the various scales to the void
correlations we investigate the alignment between the void shape and
the tidal field smoothed over a range of scales $R$.  The smoothing is
done in Fourier space using a Gaussian window function ${\hat
W}^*({\bf k};R)$:
\begin{eqnarray}
{\mathcal T}_{ij}({\bf r};R)\,=\,\qquad\qquad\qquad\qquad\qquad\qquad\qquad\qquad\qquad\ &&\  \\
{\displaystyle 3 \over \displaystyle 2}
\Omega H^2 \int \d3k\left({\displaystyle 
k_i k_j \over \displaystyle k^2}-{\displaystyle 1 \over \displaystyle 3} 
\delta_{ij}\right)\,{\hat W}^*({\bf k};R)\,{\hat \delta}({\bf k})\,
{\rm e}^{-{\rm i}{\bf k}\cdot{\bf r}}&&\, \nonumber
\end{eqnarray} 
Here, ${\hat \delta}({\bf k})$ is the Fourier amplitude of the relative density fluctuation field at wavenember $\bf k$.
 
We evaluate the alignment ${\mathcal A}_{TS}(R)$ between the void
shape ellipsoid and the local tidal field tensor ${\mathcal
T}_{ij}(R)$, Gaussian filtered on a scale $R$ at the void centers.
This is done by evaluationg, for every void, the function
${\Gamma}_{TS}(m,R)$ at the void centers:
\begin{eqnarray}
{\Gamma}_{TS}(m;R)&\,=\,&\ -\ {\displaystyle \sum\nolimits_{i,j} 
{\tilde {\mathcal S}}_{m,ij}\,{\mathcal T}_{ij}({\bf r}_m;R) 
\over \displaystyle {\tilde {\mathcal S}}_m\,\,{\mathcal T}({\bf r}_m;R)}
\end{eqnarray}
where ${\mathcal T}({\bf r}_m;R)$ is the norm of the tidal tensor ${\mathcal T}_{ij}({\bf r_m})$ filtered on a scale $R$. The void-tidal alignment ${\mathcal A}_{TS}(R)$ at a scale $R$ is then the ensemble average 
\begin{eqnarray}
{\mathcal A}_{TS}(R)&\,=\,&\langle\,\Gamma_{TS}(R)\,\rangle\,.
\end{eqnarray}
which we determine simply by averaging $\Gamma_{TS}(m,R)$ over the complete sample of voids. 
 
\begin{figure}
     \mbox{\hskip -0.9truecm\includegraphics[width=0.54\textwidth]{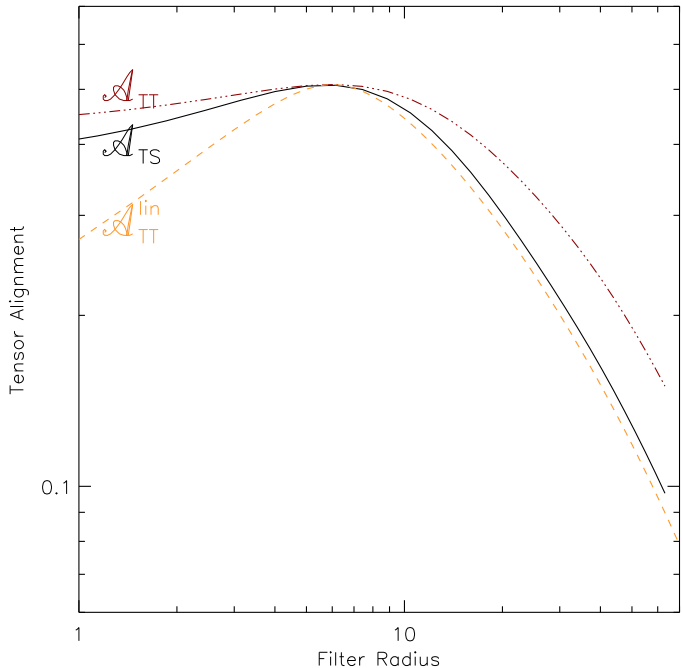}}
     \caption{Solid line: the local alignment ${\mathcal A}_{TS}(R)$ of voids with the tidal field, 
       as a function of the smoothing radius $R$ of the tidal field. Dashed line: alignment 
       ${\mathcal A}_{TT}^{lin}(R_s,R)$ between tidal field on scale $R_s=6h^{-1}\hbox{\rm Mpc}$ and 
       tidal field on scale $R$ (see text). The relation is based on the prediction for a Gaussian random field 
       with a $\Lambda$CDM spectrum. Dot-dashed line: tide-tide alignment ${\mathcal A}_{TT}(R_s,R)$ 
       measured from the N-body simulation described in the text.} 
     \label{fig:cor2}
\end{figure}
 
\subsection{Void-Tidal Field alignment: results }
A visual impression of the strong relation between the void's shape
and orientation and the tidal field is presented in the lower
righthand panel of fig.~\ref{fig:slices}. The tidal field
configuration is depicted by means of (red-coloured) tidal bars. These
bars represent the compressional component of the tidal force field in
the slice plane, and have a size proportional to its strength and are
directed along the corresponding tidal axis. The bars are superimposed
on the pattern of black solid watershed void boundaries, whose
orientation is emphasized by means of a bar directed along the
projection of their main axis.
 
The compressional tidal forces tend to be directed perpendicular to
the main axis of the void. This is most clearly in regions where the
forces are strongest and most coherent. In the vicinity of great
clusters the voids point towards these mass concentrations, stretched
by the cluster tides. The voids that line up along filamentary
structures, marked by coherent tidal forces along their ridge, are
mostly oriented along the filament axis and perpendicular to the local
tidal compression in these region.  The alignment of small voids along
the diagonal running from the upper left to the bottom right is
particularly striking.
 
To quantify and trace the tidal origin of the alignment we have
plotted in figure~\ref{fig:cor2} the local shape-tide alignment
function $\Gamma_{TS}$ versus the smoothing radius $R$. By smoothing
over a scale $R$ we suppress the contribution by mass concentrations
and structures with a scale smaller than $R$. Likewise, we have
evaluated the tide-tide alignment ${\mathcal A}_{TT}(R_1,R)$ as a
function of the same variable filtering scale $R$. The filter scale
$R_1$ (eqn.~\ref{eq:att}) is fixed at a value of $R_1\approx
6h^{-1}\hbox{\rm Mpc}$, slightly smaller than the mean void size. At
this scale we expect a strong influence of the local shape on the
tidal field, so that this quantity is expected to provide additional
information on the dynamical influence of the large-scale tidal field
on the dynamics of the void. Note that this implies there to be a
slight offset between the actual mean void size, $R_v \approx
8.5h^{-1}\hbox{\rm Mpc}$ and the peak scale of ${\mathcal
A}_{TS}$. The latter is biased towards the more strongly clustered
small voids, whose alignment is often due to one coherent nearby
large-scale structure, often a filament or cluster.
 
\medskip
\noindent The following observations can be made:

\begin{list}{}
	{
	\setlength{\itemindent}{-10pt}
	\setlength{\leftmargin}{10pt}
	}
	
\item[$\bullet$] The alignment is strong over the whole range of smoothing radii out to $R \approx 20-30h^{-1}\hbox{\rm Mpc}$ (${\mathcal A}_{TS}\approx 0.25$).
 
\item[$\bullet$] The alignment peaks at a scale of $R \approx 6 h^{-1} \hbox{\rm Mpc}$. This scale is very close to the average void size (see fig.~\ref{fig:shape}), and also close to the scale of nonlinearity. The latter is not a coincidence: the identifiable voids probe the linear-nonlinear transition scale. 
 
\item[$\bullet$] There is still a remarkably strong alignment signal at radii larger than $R>20h^{-1}\hbox{\rm Mpc}$ (where ${\mathcal A}_{TS}\approx 0.3$). It forms a strong argument for the substantial role of the large scale tidal field in aligning the voids. If there were no external large-scale contribution the alignment $\Gamma_{TS}$ would have disappeared for smoothing radii $R$ in excess of the average void scale.  
 
\item[$\bullet$] The fact that the alignment $\Gamma_{TS}$  at the peak scale $R \approx 6 h^{-1} \hbox{\rm Mpc}$ is less than twice that of at $R \approx 25 h^{-1} \hbox{\rm Mpc}$ shows that local contributions, in particular due to the voids' own ellipticity, are only partially responsible for the shape-tidal field alignment. 
\end{list}
We have also analyzed higher resolution simulations. The results we
found here carry over to the higher resolved simulations. The smaller
box of the latter, however, contain fewer voids and there is a smaller
range over which the influence of tidal fields can be included. The
results do remain significant but are marked by more noise.
 
\subsection{Tidal Connections}
We have established a strong correlation between void orientation and
the large scale tidal field.  We now compare this with the
self-alignment of the large scale tidal field that would be expected
on the basis of linear perturbation theory.  The dashed curve in
fig.~\ref{fig:cor2} is the linear theory tide-tide alignment
${\mathcal A}_{TT}^{lin}(R_s,R)$, with $R_s=6h^{-1}\hbox{\rm Mpc}$.
On scales greater than the void scale the void-tidal field alignment
follows the tidal field alignment. On smaller scales, however, we
start to see a marked deviation: the alignments are stronger than
expected from the initial linear field. This may be because of
nonlinear effects on small scales: the tidal field on the smaller
scales where the density fluctuaions are nonlinear is even strong
enough to cause the destruction of small voids near the filaments of
the cosmic web where there is a strong anisotropic force field
\citep[see][]{weysheth2004}.
 
We also measured the tide-tide alignment ${\mathcal A}_{TT}(R_v,R)$
for the tidal field of our $\Lambda$CDM cosmology where any nonlinear
effects that are present are accounted for.  The dot-dashed curve is
the result (note that the amplitude has been scaled so that the
amplitudes of both ${\mathcal A}_{TT}$ curves match the peak of the
${\mathcal A}_{TS}$ curve).  The fact that, on small scales, there is
clear agreement between the void-tide alignment and the tide-tide
alignment underlines the important role of tidal forces throughout the
formation history of structures in the cosmic web.

As an incidental remark, the existence of the long range tide - void
alignment tells us that the void anisotropy is not a mere artefact of
the WVF method chopping spherical voids in two (which would of course
lead to non-spherical voids having correlated alignments).
 
\section{Conclusion and Discussion}
We have investigated the shapes and alignments of voids found in the
large scale matter distribution of the VLS GIF N-body simulation of
structure formation in a $\Lambda$CDM universe.  The void sample has
been identified using the Watershed Void Finder (WVF) technique.
 
\subsection{Overview of Results}
Although voids tend to be less flattened or elongated than the halos
in the dark matter distribution, they are nevertheless quite
nonspherical: they are slightly prolate with axis ratios on the order
of $c:b:a\approx 0.5:0.7:1$.  Two important factors contribute to this
flattening.  Internally, voids tend to become more spherical as they
expand \citep{icke1984}. However, they will never be able to reach
perfect sphericity before meeting up with surrounding structures.
Furthermore, external tidal forces will contribute substantially to
the anisotropic development of the voids. In this study we have
established the influence of the tidal fields as being an important
agent in the dynamical evolution of voids.  Indeed, our study provides
strong confirmation that tidal forces dominate the shaping of the
Cosmic Web, as emphasized long ago by \cite{bondweb1996}.
 
We have also investigated the relative orientation of the voids. The
orientation of voids appears to be strongly correlated with alignments
spanning distances $>30h^{-1}\hbox{Mpc}$.  This is true of both the
shortest and longest axes of the void shape.  This coherence is quite
conspicuous in plots of the density distribution. To test for such
void alignments in observational data a follow-up study will apply the
WVF formalism to galaxy redshift surveys.
 
We find an intimate link between the cosmic tidal field and the void
orientations. Over a range of scales we find a strong alignment of the
voids with the tidal field arising from the smoothed density
distribution. Given that the alignment correlations remain significant
on scales considerably exceeding twice the typical void size, our
results show that the long range external field is responsible for the
alignment of the voids. This confirms the view that the large scale
tidal force field is the main agent for the spatial organization of
the Cosmic Web.
 
\subsection{Insights}
We have argued that the large scale tidal field is not only the main
agent responsible for the shapes of the voids but also for the
resulting alignments of the voids. Locally, the orientation of a void
turns out to be strongly aligned with the tidal force field generated
by structures on scales up to at least $20-30h^{-1}\hbox{\rm Mpc}$. On
scales comparable to the average void size we also found indications
of nonlinear effects, with the void's orientation reacting to the
small-scale nonlinear influences.
 
This conclusion agrees with that of a similar study of halo alignments
by \cite{leespringel2007}. However, the alignment of halos tends to be
strongly attenuated by local nonlinear effects
\citep[e.g.][]{haarwey1993, aragon2007} rendering the final halo
alignment signal weaker than that for voids.
 
One aspect we have not yet adressed in detail is the entanglement of
initial conditions and the influence of the tidal field. On large
scales, where the density fluctuations are in the linear regime, these
are different aspects of the inhomogeneous matter distribution: if one
is known the other can be determined. In this respect the indication
of nonlinear effects in the tidal-void alignment ${\mathcal A}_{TS}$
(fig.~\ref{fig:cor2}) at small scales is interesting. They may very
well be an expression of the strong influence of coherent overdense
filaments and dense compact clusters on the evolution of small voids
in the outer regions of large voids.
 
\cite{weysheth2004} argued that the collapse of small voids, an
essential aspect of the hierarchical evolution of voids
\citep{shethwey2004}, manifests itself in a tidally induced
anisotropic contraction of small underdensities at the void
boundaries. The dynamics and the fate of these small underdensities is
often decisively influenced by external anisotropic forces
\citep{weybabul1996}. While such forces may affect the nonlinear
evolution of halos and clusters to a considerable extent
\citep{bondmyers1996,sheth2001}, they do not go as far as deciding
their fate.
 
This work is part of a study of the evolution of voids under the
influence of external influences. A proper framework for following the
gradual hierarchical buildup of void regions under the influence of
external tidal forces is provided by the void-patch description
\citep{platen2008, bondweb1996}. To some extent the void-patch
formalism is a more accurate approximation of reality than for the
equivalent peak-patch: because of their approximate uniformity void
evolution is accurately described by the ellipsoidal model.
 
A perhaps even more complex issue is that of the identity of the outer
regions of voids, the regions where the void merges into the
surrounding matter distribution. We have argued that this will play an
important role in determining the shape of voids. However, a proper
definition of a void boundary does not (yet) exist. Arguably the WVF
has proven to represent a major advance along these lines and we
should soon be able to be more definite on the issue.
 
Having established the external influences on a void's evolution, we
may try to understand the sensitivity of the void population to the
underlying cosmology. This depends to a considerable degree on the
differences in scale dependence of the tidal force
fields. \cite{leepark2007} did find a considerable influence. This
will open the door to the use of voids in determining the global
cosmology.
 
\section*{Acknowledgements}
We gratefully ackowledge the use of the Virgo Consortium simulation. We thank Miguel Arag\'on-Calvo for useful discussions on various technical aspects of this paper.

\label{lastpage}
 
\end{document}